\documentclass{ap-jnmp}
\usepackage{multicol}
\usepackage{color}
\usepackage{graphicx}

\usepackage[english]{babel}
 
\title{A simple-looking relative of the Novikov, Hirota-Satsuma and Sawada-Kotera equations} 
\author{Alexander G. Rasin}
\address{
Department of Mathematics, Ariel University,\\
 Ariel 40700, Israel \\
\email{rasin@ariel.ac.il}}

\author{Jeremy Schiff}
\address{
Department of Mathematics, Bar-Ilan University,\\
 Ramat Gan, 52900, Israel \\
\email{schiff@math.biu.ac.il}}

\begin{document}
\maketitle
\thispagestyle{empty}

\vphantom{\vbox{%
\begin{history}
\received{(Day Month Year)}
\revised{(Day Month Year)}
\accepted{(Day Month Year)}
%\comby{(xxxxxxxxx)}
\end{history}
}}

\begin{abstract}
  We study the simple-looking scalar integrable equation $f_{xxt} - 3(f_x f_t - 1) = 0$, which is related (in different ways) to
  the Novikov, Hirota-Satsuma and Sawada-Kotera equations. For this equation we present a Lax pair, a B\"acklund transformation,
  soliton and merging soliton solutions (some exhibiting instabilities),  two infinite hierarchies of conservation laws,
  an infinite hierarchy of continuous symmetries, a Painlev\'e series, a scaling reduction to a third order ODE and its
  Painlev\'e series, and the Hirota form (giving further multisoliton solutions). 
\end{abstract}

\ccode{2000 Mathematics Subject Classification: 35Q58, 37K05, 37K10, 37K35, 37K40, 37K45, 70G65}

\section{Introduction}

Early in the history  of integrable systems it was found that there are remarkable transformations
connecting different equations \cite{Mi12}. These ``hidden connections'' abound, and
complicate classification attempts, see for example \cite{MSY}. But, in general, discovery of
such a connection is very useful, as typically most  of the integrability  properties of
an equation can be mapped to other related equations. Thus, for example, more research focuses on the KdV equation
than on the (defocusing) MKdV equation, and on the NLS equation rather than the many systems related to it
by Miura maps (see Figure 1 in \cite{Sch1}). Naturally, the focus tends to fall on the equation in the
class which is simplest to write down.

One of the simple, archetypal equations in integrable systems theory is the Camassa-Holm equation, the first
integrable equation to be discovered with peakon solutions \cite{CH,CHH}. This was followed by the
Depasperis-Procesi equation \cite{DP,DHH} and by the Novikov equation \cite{No1,HW0}
\begin{align}
&m_t+3uu_xm+u^2m_x=0,&\label{um1}\\
&m=u-u_{xx}.\label{um2} &  
\end{align}
In  \cite{RS7} we announced that the Novikov equation is related, by a chain of transformations, to the particularly
simple equation 
\begin{equation}
f_{xxt}-3(f_xf_t-1)=0\ ,   \label{aN} 
\end{equation}
which we called the associated Novikov equation (aN). We believe that \cite{RS7} is the first time (\ref{aN})
was written down explicitly as a single equation. However, it is trivial to derive from equations (2.6a) and
(2.8) in \cite{ma2013} by introducing a potential (writing $W=-F_\tau$, $U=3F_y$) followed by a rescaling
(and changes of names of the variables).  See also \cite{WLL0}, equation (9). Matsuno derived a
different scalar equation from equations (2.6a) and (2.8) in \cite{ma2013}. In our notation, note that we can solve
(\ref{aN}) to find $f_t$ in terms of $g=f_x$. Differentiating this with respect to $x$ and rearranging gives 
\begin{equation}
 gg_{xxt}  - g_xg_{xt} - 3 g^2 g_t  - 3g_x = 0  \ ,  \label{matsunoeq} 
\end{equation}  
c.f. (2.11) in \cite{ma2013}.
{{Note also that by rescaling $t$ it is possible to replace the constant term in (\ref{aN})  by any non-zero constant.
    The equation with vanishing constant term, i.e.
\begin{equation}
f_{xxt}-3 f_xf_t =0\ ,   \label{aN2} 
\end{equation}
was studied in \cite{MR1109192}, and we will discuss this equation further at the end of this introduction.}}    

Since the derivation of the various equations in \cite{ma2013} from the Novikov equation is clear, we do
not give the full derviation of aN from Novikov here. Proceeding from (\ref{aN}), however, make the substitution
$$ f(x,t) = h(x,t)  + \beta x + \frac{t}{\beta}\ ,    $$
where $\beta\not=0$ is a constant, to obtain
$$  h_{xxt} - 3 h_xh_t + \beta h_t + \frac{h_x}{\beta} = 0 \ .  $$ 
Differentiating with respect to $x$ and  writing $h=r_x$, so $r=-\int_x^\infty h(y,t)\ dy$, we obtain
\begin{equation}
r_{xxt} + 3 r_{x} \int_x^\infty r_t(y,t)  dy  - 3rr_t    + \beta r_t + \frac{r_x}{\beta} = 0 \ .  \label{hirsat} 
\end{equation}  
Modulo rescalings, this is equation (2) in \cite{HS1}, known as the Hirota-Satsuma equation. 
In \cite{ma2013}, Matsuno established the connection between his equation (\ref{matsunoeq}) and the
Hirota-Satsuma equation. 

Thus we see that aN is related to the Novikov equation (\ref{um1})-(\ref{um2}), the Matsuno equation (\ref{matsunoeq}) 
and the Hirota-Satsuma equation (\ref{hirsat}). We note, though, that aN is rather more simple in form than any
of these other equations, and thus is surely the natural first object for study. 
In fact there is a further equation to which aN is related, though this time it is not by a sequence
of transformations. We recall that $\eta$ is an infinitesimal symmetry for (\ref{aN}) if, as a consequence of (\ref{aN}),
$f+\epsilon \eta$ is a solution of (\ref{aN}) up to linear order in $\epsilon$. That is, if
$$   \eta_{xxt} - 3 f_x\eta_t - 3 f_t\eta_x = 0\ .  $$ 
A direct calculation shows that  $ \eta = -\frac1{15} f_{xxxxx} + f_xf_{xxx} - f_x^3  $ 
satisfies the necessary condition. It follows that the flow
\begin{equation}
  f_s = \frac1{15} f_{xxxxx} -f_xf_{xxx}  + f_x^3 \label{pSK}
\end{equation}  
(now $f$ is being considered as a function of three variables, $x,t$ and $s$) is consistent with the flow defined by
aN. This last equation is a potential form of the Sawada-Kotera equation (SK), viz. equation (44) in  \cite{SW0}. 
The link between aN and potential SK is not that there are  transformations from one to the other, but they define consistent
flows, i.e. they belong to the same {\em hierarchy}. Although the link is less direct, integrable equations in the same hierarchy also
share properties, such as conservation laws and soliton solutions. The fact that the Novkiov equation is related to the SK hierarchy was
already established in \cite{HW0}, see also \cite{KLO}.  Although in this case  aN has a complication not 
in the potential SK flow, in that  in equation (\ref{pSK}) only a single $t$ derivative term appears, aN remains superficially
simpler, so we would argue that aN is the right equation to study first. 

We have thus seen in this introduction how aN, equation (\ref{aN}), is related to numerous equations of interest, but
is, at least superficially, the simplest of all. The aim of this paper is to present the basic properties of
aN, without reference to any other equation, with the intention that ultimately, aN can be used as a base for
study of the other equations. In section 2 we present the Lax pair and B\"acklund transformation (BT) for aN. 
{{The Lax pair, which we derived from the Lax pair for the Novikov equation given in \cite{HW0}, 
is identical to that given for (\ref{aN2}) in \cite{MR1109192}.}}   In section 3
we present the basic soliton solutions of aN, and also solutions describing the ``merger'' 
of two solitons, obtained using the BT. However, the BT also gives rise to a solution showing that at least some of the basic soliton
solutions are {\em unstable}. 
This limits the possible physical relevance of aN, but on the other hand provides a simple analytic example
of the phenomenon of soliton instability, the possibility of which is often overlooked.  In section 4 we discuss conservation laws,
giving two infinite hierarchies. In section 5 we consider symmetries, but only succeed to give one infinite hierarhcy.
We believe a second hierarchy should exist (in parallel to what we have found for conservation laws), but leave
this as an open problem. We show the existence of a second infinite hierarchy in the case of the Camassa-Holm equation
to demonstrate that this is feasible. In section 6 we show the Painlev\'e property for aN. In section 7 we discuss
a scaling reduction to a third order ODE and give its Painlev\'e series and the simplest solution. In section 8 
we give the Hirota form of aN and a formula for multisoliton solutions. In section 9 we conclude. 

{{Before leaving this introduction, we briefly discuss equation (\ref{aN2})
that was studied in \cite{MR1109192}.
As mentioned above, by rescaling $t$ it is possible to replace the constant in (\ref{aN}) by any nonzero constant,
and thus it is possible to derive properties of the equation (\ref{aN2}) from those of the equation (\ref{aN})
by a limit process. In this way, for example,
it is possible to find (several families of) soliton solutions of the (\ref{aN2}).
For a reason that will become clear in the next section, there is a single Lax pair for the equation, irrespective of the
value of the constant term. However, in general we are not aware of a process to derive properties of (\ref{aN})
from those of (\ref{aN2}) and we believe the two equations are nonequivalent.
Furthermore, it is (\ref{aN}) that arises in the context of the Novikov equation. 
Thus in this paper we study  (\ref{aN}). 
}}

\section{Lax Pair and B\"acklund transformation}

The Lax pair for aN is 
\begin{align}
\phi_{xxx}&=3f_x\phi_x+\theta\phi,&\label{psixxx}\\
\phi_t&=\frac{1}{\theta}(f_t\phi_{xx}-f_{xt}\phi_x).\label{psit}&
\end{align}
As mentioned above, this Lax pair can be derived, 
through the necessary transformations, from the Lax pair for the Novikov equation given in \cite{HW0}.
{{
    In fact the consistency condition for the Lax pair is not exactly equation (\ref{aN}), but rather
    its $x$-derivative. This explains why it  also provides a Lax pair for the equation (\ref{aN2}), as studied
    in  \cite{MR1109192}, and indeed it coincides with the Lax pair given there.}}    
The first equation of the Lax pair coincides with the first equation of the Lax pair for SK \cite{Ka1}.  

The B\"acklund transformation (BT) for aN is  
\begin{equation}
f\rightarrow f-2v,\label{BT}
\end{equation}
where $v$ satisfies the following system:
\begin{align}
v_{xx}&=-v^3-3vv_x+3f_xv+\theta,&\label{vf1}\\
v_t&=f_t-\frac{f_tv^3+f_tvv_x-f_{tx}v^2-3v}{\theta}.\label{vf2}&
\end{align}
The second equation can be rewritten 
\begin{equation}
v_t=\frac{1}{\theta}(f_tv_{x}-f_{xt}v+f_{t}v^2)_x.  \label{cl}
\end{equation}
The equation (\ref{vf1}) has appeared before in \cite{Nu1, SAMS} as part of the BT for SK. 
The BT (\ref{vf1},\ref{vf2}) is connected to the Lax pair (\ref{psixxx},\ref{psit}) 
via the substitution
\begin{equation}
  v=\frac{\phi_x}{\phi}.  \label{LP2BT} 
\end{equation}  
By the relevant transformations, this BT is equivalent to the BT given by Yadong Shang \cite{SY1}  for the Hirota-Satsuma  
equation. 

\section{Soliton and merging soliton solutions}

Using the standard procedure for finding travelling wave solutions shows that aN has soliton solutions
$$ f =  \beta x + \frac{t}{\beta}  -2 B  \tanh B(x-ct) \ ,  \qquad B^2 = \frac34 \left( \beta -\frac1{c\beta}  \right)\ .    $$
We call this a soliton solution as the associated profile $g=f_x$ of the Matsuno equation (\ref{matsunoeq}) has 
a ${\rm sech}^2$ profile, on the nonzero background $\beta$ (in fact the profile is $-{\rm sech}^2$, so it might more correctly
be called an antisoliton). Note that here $\beta\not=0$ and $c$, the wave speed, are arbitrary, subject to the constraint
$$      \beta > \frac1{c\beta}\ .    $$
So if $\beta$ is negative, then $0<c<\frac1{\beta^2}$, and if $\beta$ is positive then either $c$ is negative or
$c>\frac1{\beta^2}$.  In the solution it is possible to replace $\tanh$ by $\coth$  to get a singular soliton solution.  

Application of the BT (\ref{BT}),(\ref{vf1}),(\ref{vf2}) to the ``trivial'' solution 
$$   f =  \beta x + \frac{t}{\beta}      $$
gives the solution  
\begin{equation}
f =  \beta x + \frac{t}{\beta} -2\frac{ C_1\lambda_1 e^{\lambda_1 x+\frac{\lambda_1^2t}{\beta\theta}}
                                      + C_2\lambda_2 e^{\lambda_2 x+\frac{\lambda_2^2t}{\beta\theta}}
                                      + C_3\lambda_3 e^{\lambda_3 x+\frac{\lambda_3^2t}{\beta\theta}}}
                                      { C_1 e^{\lambda_1 x+\frac{\lambda_1^2t}{\beta\theta}}
                                      + C_2 e^{\lambda_2 x+\frac{\lambda_2^2t}{\beta\theta}}
                                      + C_3 e^{\lambda_3 x+\frac{\lambda_3^2t}{\beta\theta}}}.       \label{1BT}
\end{equation}
where   $C_1,C_2,C_3$ are constants and 
$\lambda_1,\lambda_2,\lambda_3$ are the roots of  the cubic equation  $\lambda^3 = 3 \beta\lambda + \theta$.  
The analysis of these solutions proceeds in a similar manner as in \cite{RS6,RS7}.
We restrict the analysis to the case $4\beta^3 > \theta^2$, so the roots $\lambda_1,\lambda_2,\lambda_3$ are real and distinct;
note, in particular, that this implies $\beta>0$. 
When one of the constants $C_i$ vanishes, say $C_3$, we recover a travelling wave solution with speed $c=\frac{\lambda_3}{\beta\theta}$,
which is a soliton if $C_1,C_2$ have the same sign and a singular soliton if they have different signs.  It is possible to check that for a suitable
choice of $\theta$ this gives all possible speeds $c$ with $c<0$ or  $c>\frac1{\beta^2}$. 

Continuing now  to the case that all the three constants $C_1,C_2,C_3$ are nonzero,
if all of them have the same sign then evidently the solution (\ref{1BT}) will be 
nonsingular, but if there are differing signs then we expect a singularity. 
Without loss of generality assume $\lambda_1<\lambda_2<\lambda_3$. Then it can be checked that for $\theta>0$ the solution
with $C_1,C_2,C_3$ all of the same sign 
describes solitons of speeds  $\frac{\lambda_1}{\beta\theta}$  and $\frac{\lambda_3}{\beta\theta}$  merging to give a
soliton of speed  $\frac{\lambda_2}{\beta\theta}$,
while for $\theta<0$ the opposite happens, with  the soliton of speed  $\frac{\lambda_2}{\beta\theta}$ splitting into two
solitons of speeds $\frac{\lambda_1}{\beta\theta}$  and $\frac{\lambda_3}{\beta\theta}$.
See Figure 1 (for the case $\theta>0$).   Note that $\lambda_1+\lambda_2+\lambda_3=0$
(so in the mergers there is ``conservation of speed''), and $\lambda_1\lambda_2\lambda_3 = \theta$.   
So in general  $\lambda_1<0$ and $\lambda_3>0$, while the sign of $\lambda_2$ is opposite to that of $\theta$. 

\begin{figure}
      \centerline{
        \includegraphics[width=4.5cm]{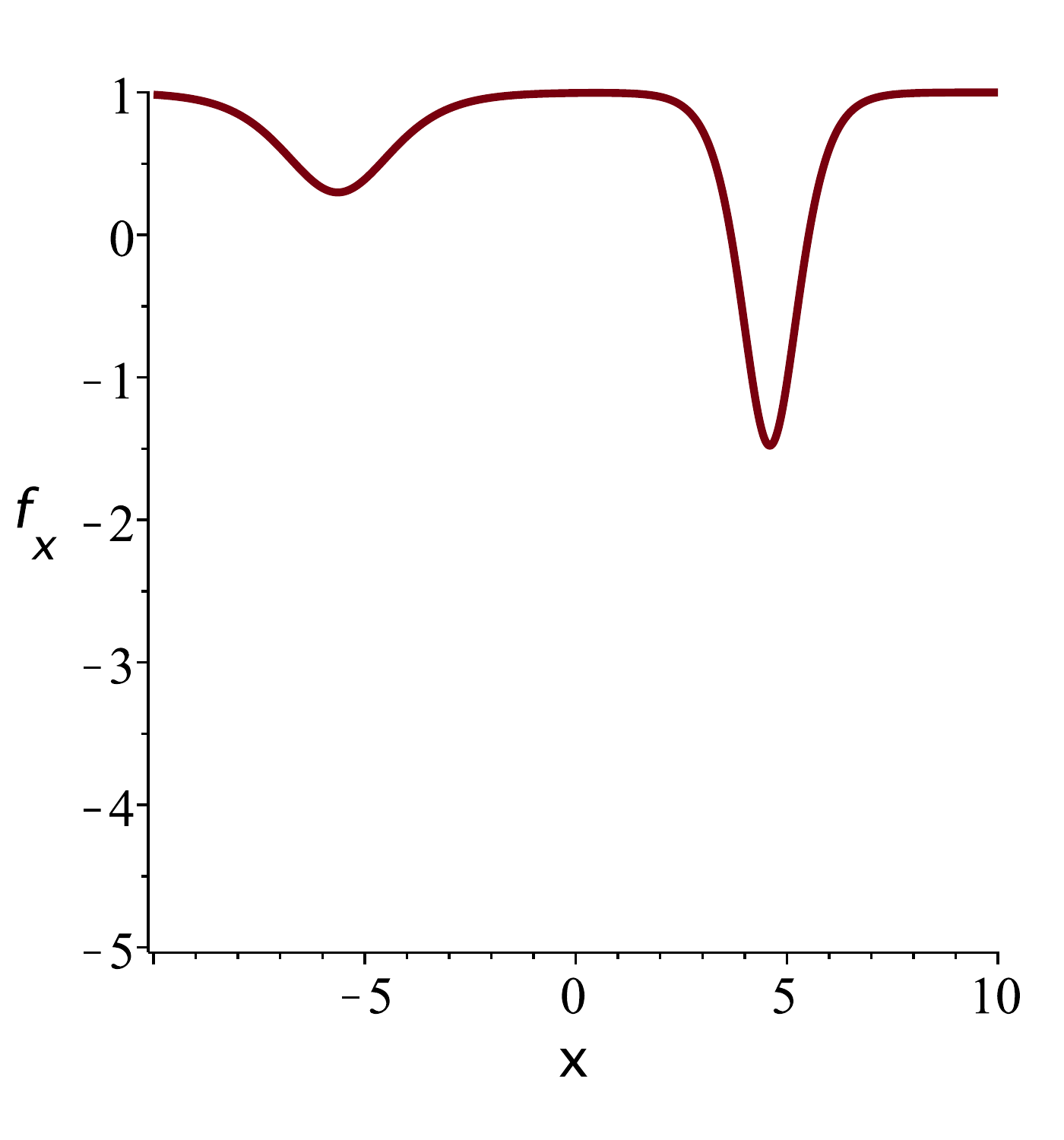} 
        \includegraphics[width=4.5cm]{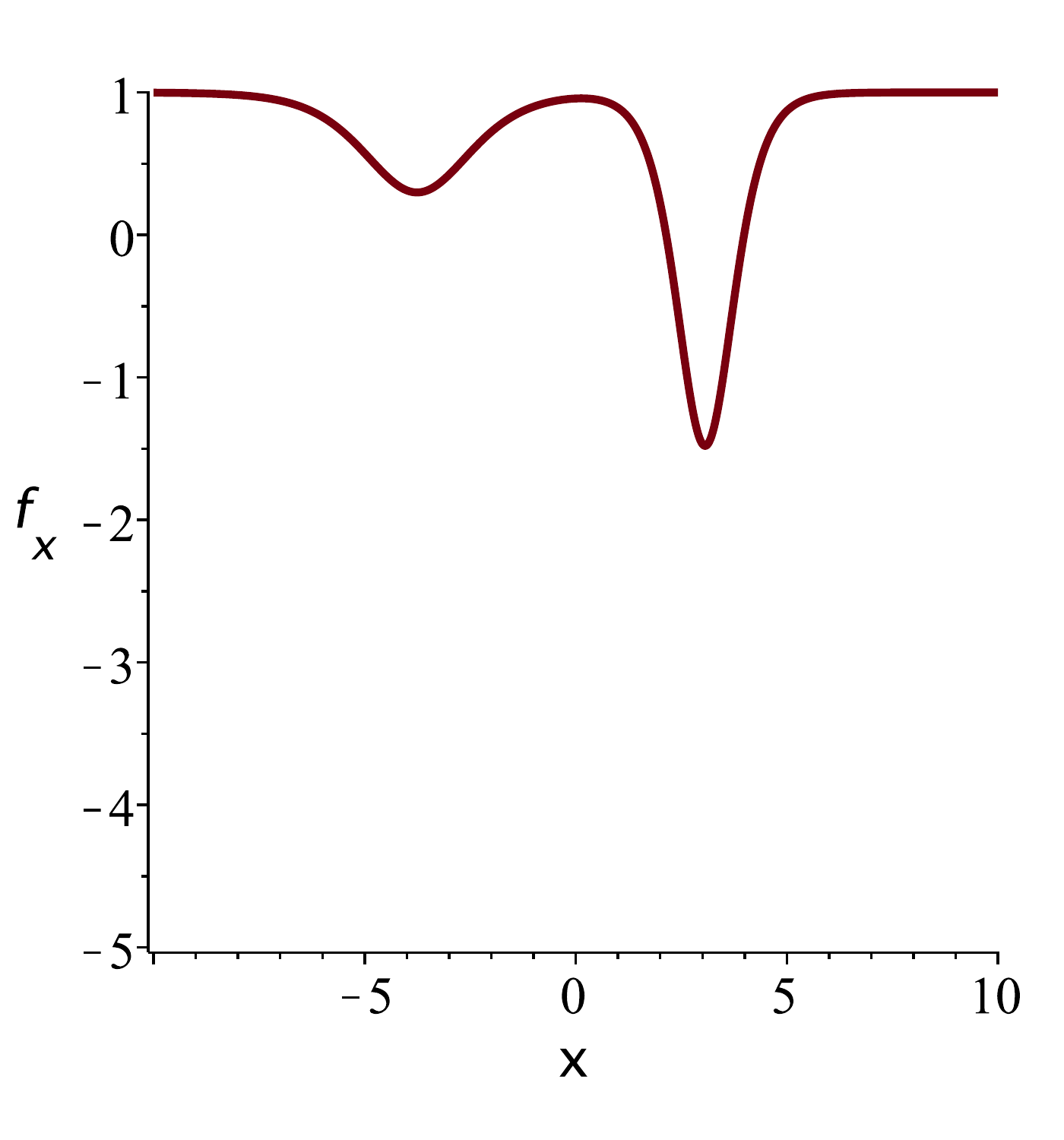} 
        \includegraphics[width=4.5cm]{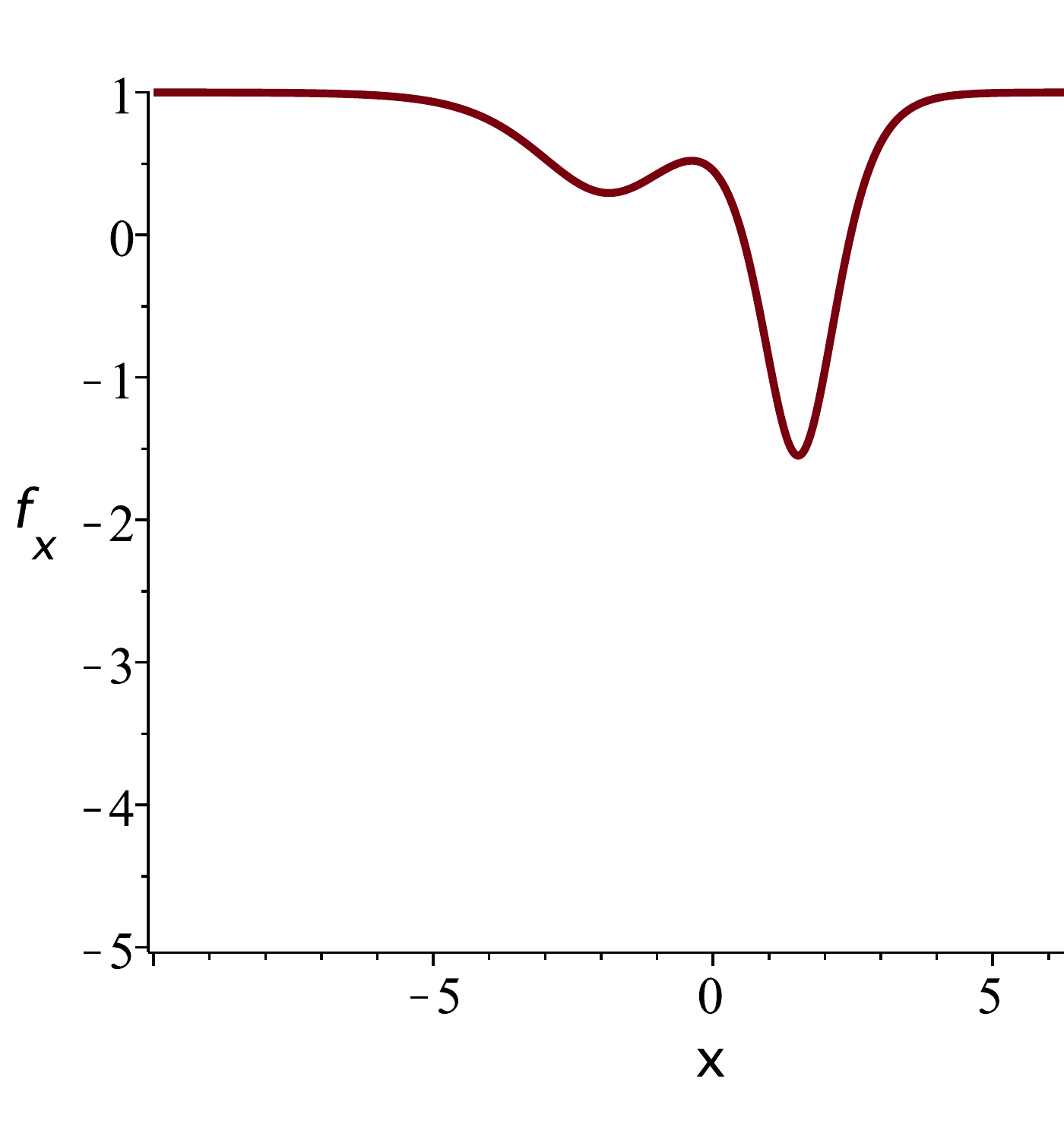} 
        }
      \centerline{
        \includegraphics[width=4.5cm]{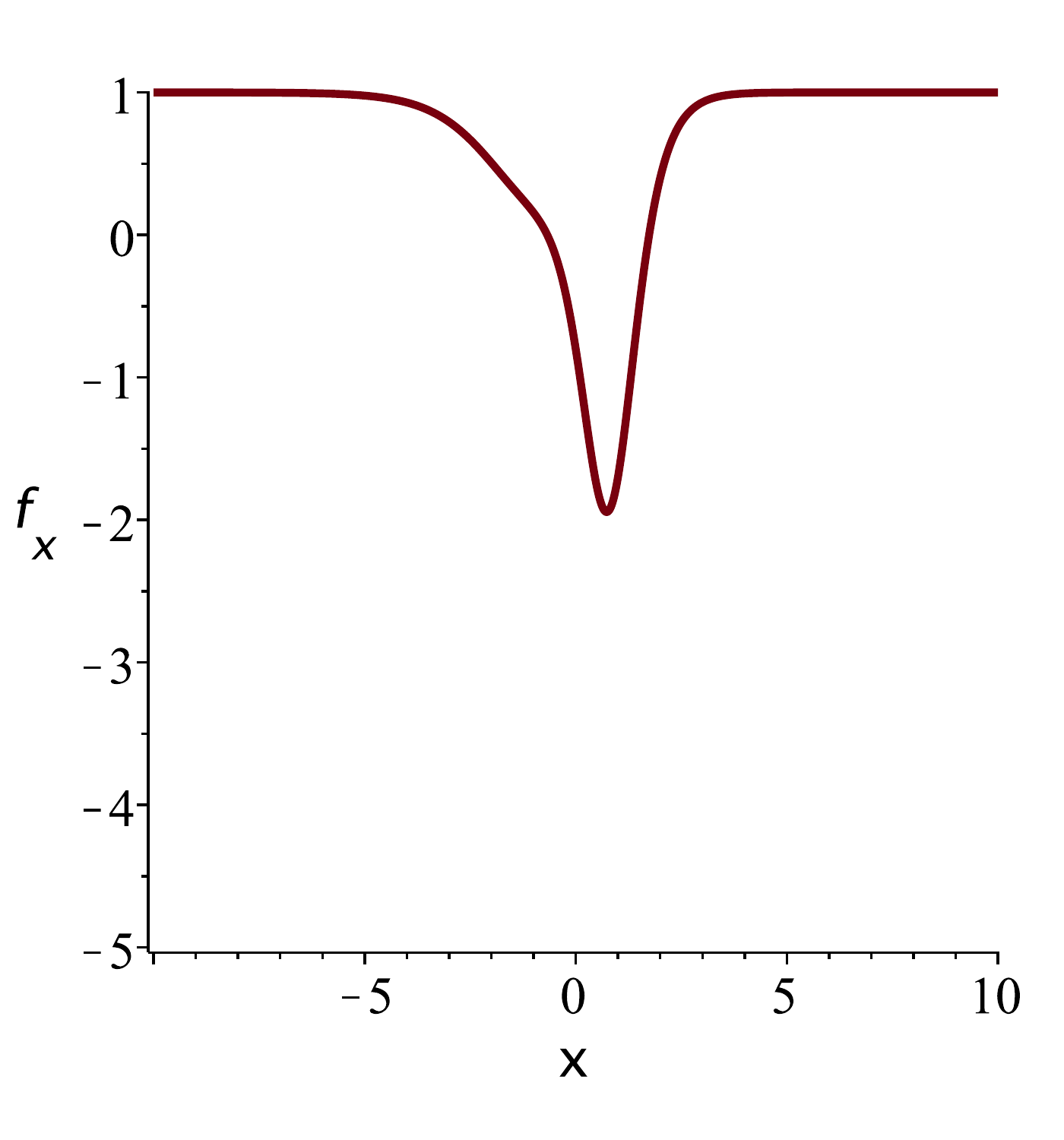} 
        \includegraphics[width=4.5cm]{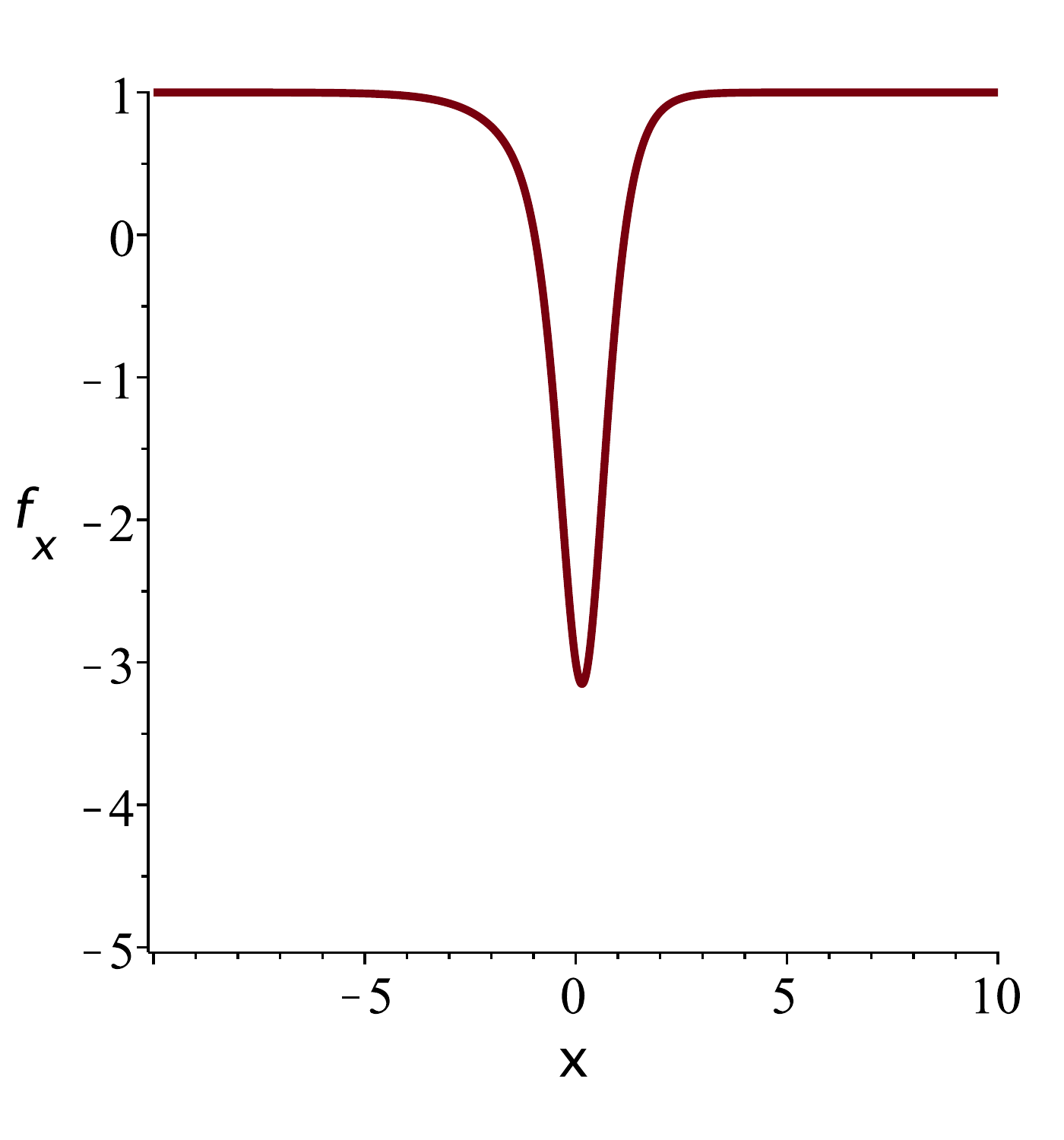} 
        \includegraphics[width=4.5cm]{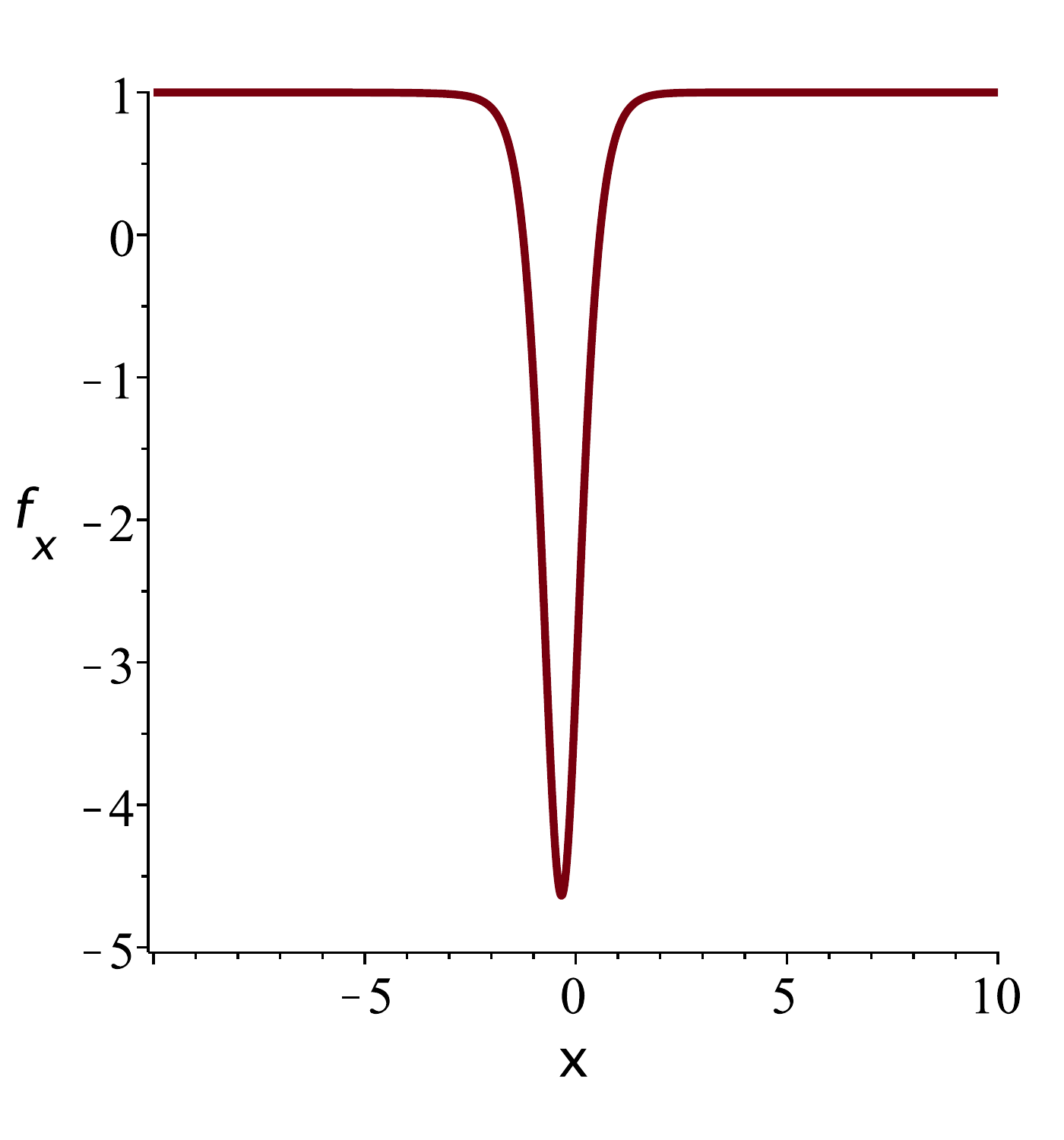} 
        }
      \caption{The merging soliton.  Parameter values are $\beta=1$ and $\theta=1$, so $\lambda_1,\lambda_2,\lambda_3 \approx -1.53,-0.35,1.88$.
        The constants $C_1,C_2,C_3$ are all taken to be $1$.  Plots of $f_x$, with $f$ given by (\ref{1BT}),  displayed for times
        $t=-3,-2,-1,-0.5,0,1$.} 
\end{figure}

When $C_1,C_2,C_3$ have differing signs, various  scenarios emerge. Figure 2 shows a case of 
expulsion of a soliton by a singular soliton. However, the most interesting case is when $\theta<0$
and the sign of $C_2$ differs from that of $C_1$ and $C_3$. In this case the solution describes the 
splitting of a left moving soliton into a pair of singular solitons, see Figure 3.  For large negative $t$
this solution is a small perturbation of the soliton solution obtained by setting $C_2=0$. However, as $t$
increases, the small perturbation grows, ultimately causing a divergence. Thus we deduce that the original soliton solution
is unstable. A brief calculation shows this only happens for solitons with $-\frac1{2\beta^2}<c<0$.
The stability of solitons for other values of $c$ (and for the case of negative $\beta$) remains open. 

\begin{figure}
  \centerline{
        \includegraphics[width=4.5cm]{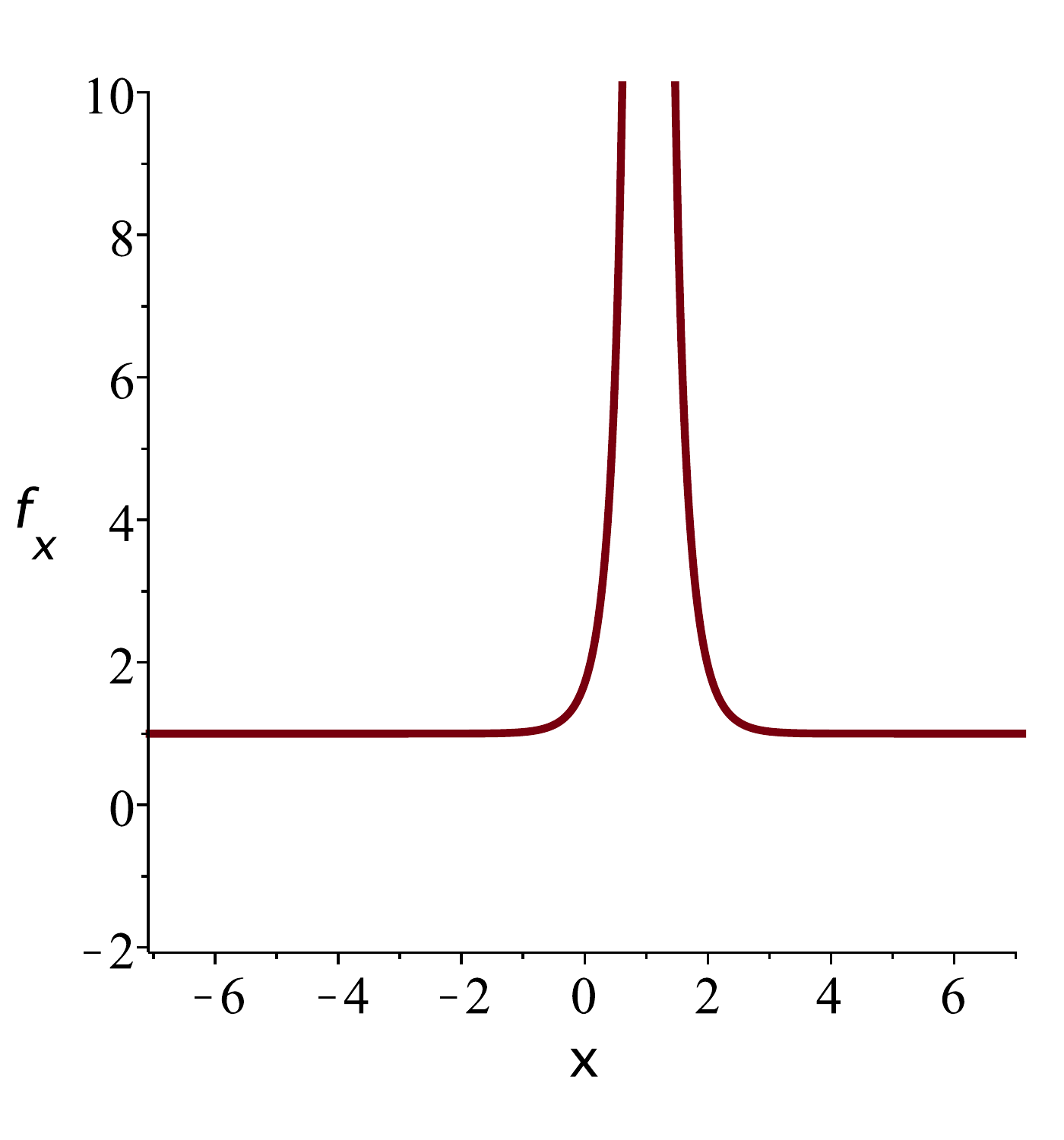} 
        \includegraphics[width=4.5cm]{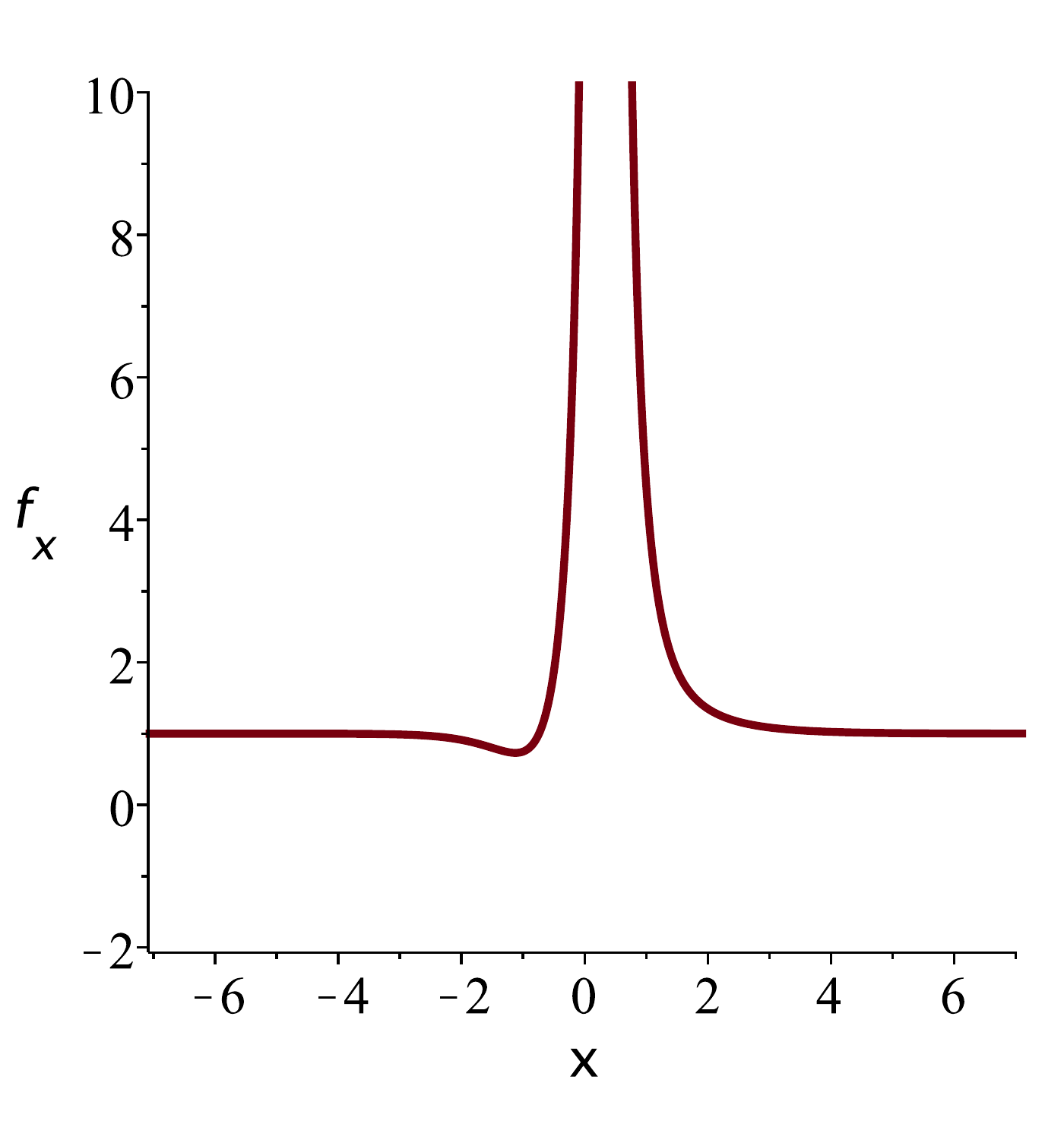} 
        \includegraphics[width=4.5cm]{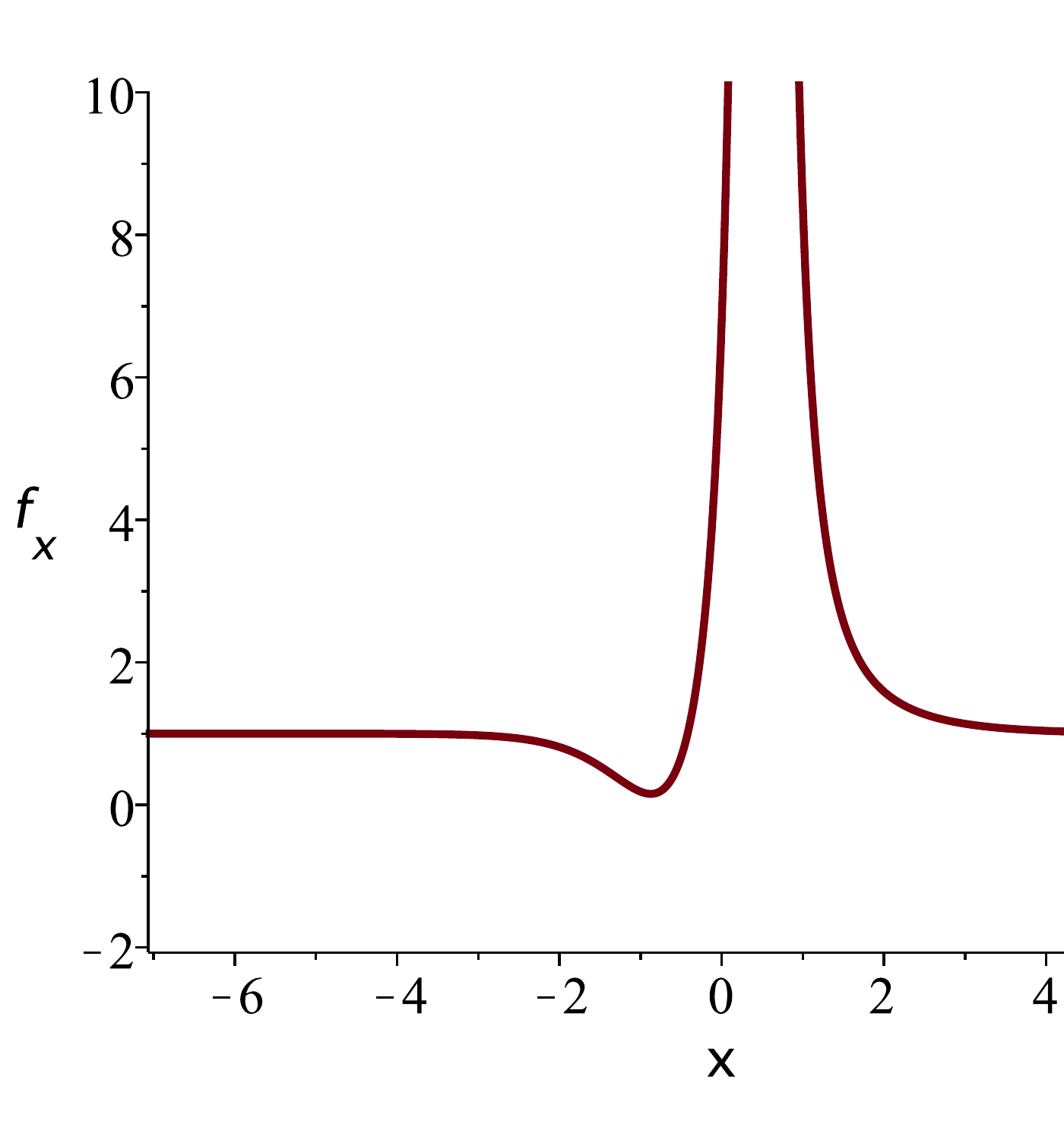} 
        }
      \centerline{
        \includegraphics[width=4.5cm]{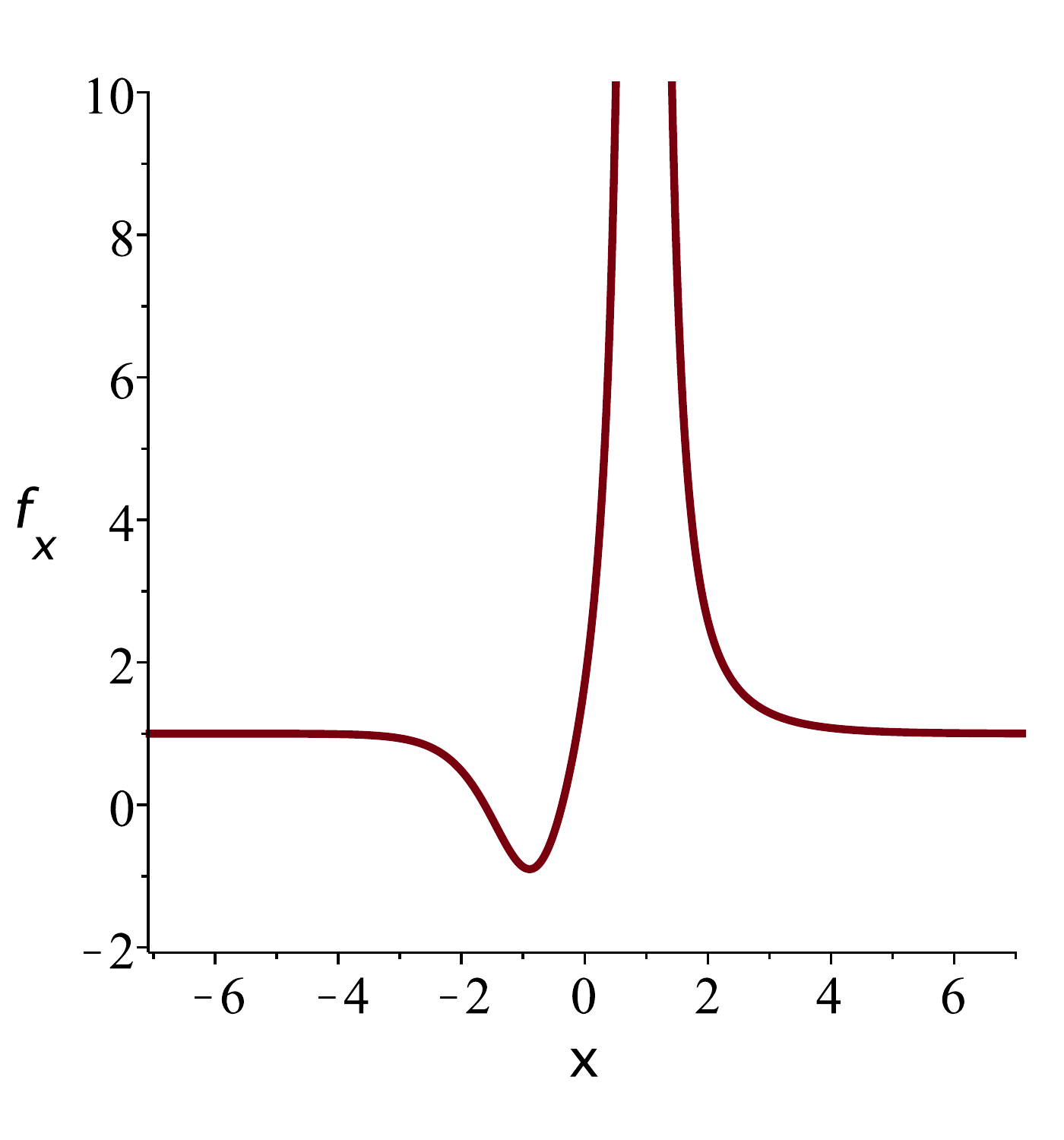} 
        \includegraphics[width=4.5cm]{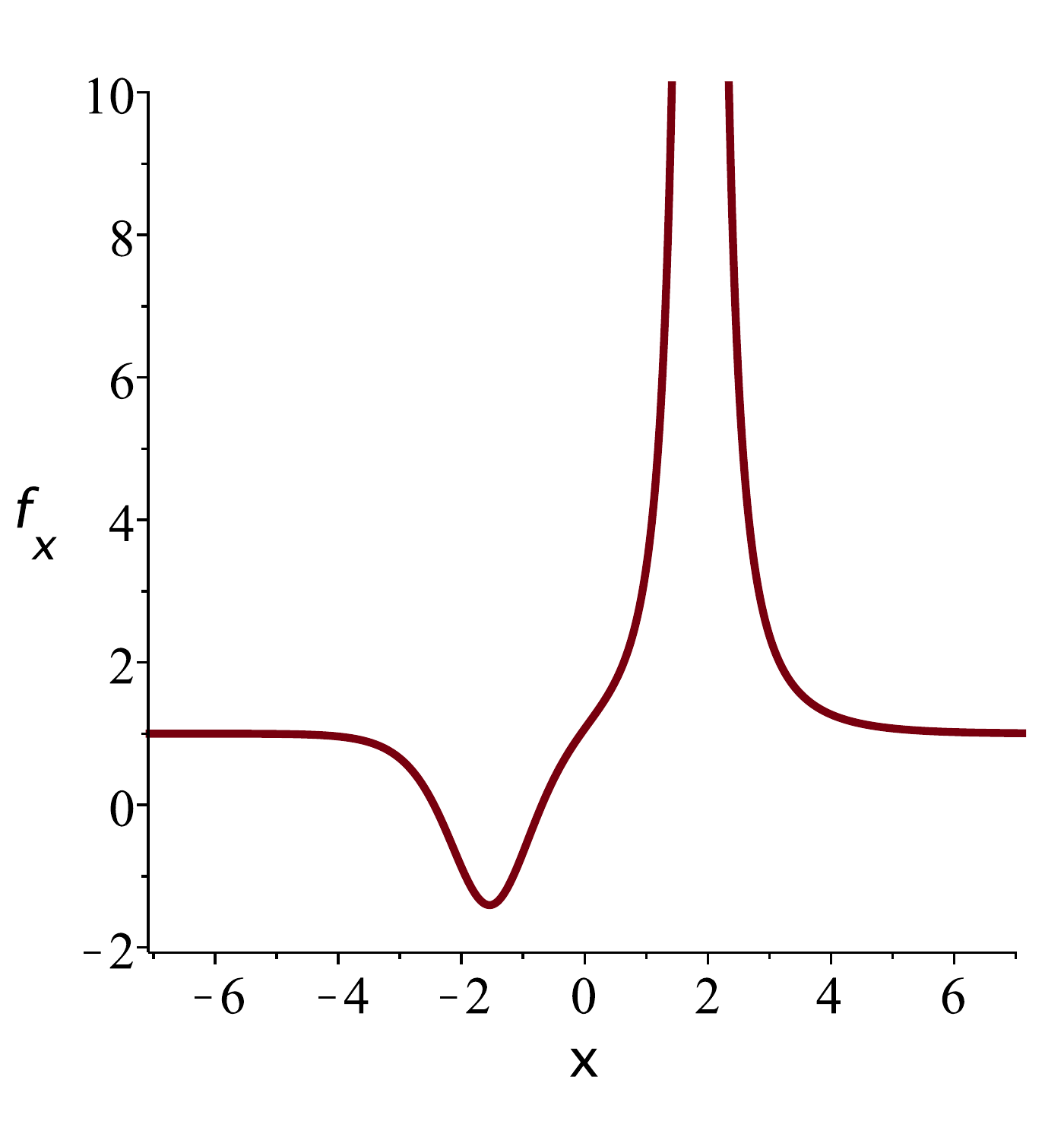} 
        \includegraphics[width=4.5cm]{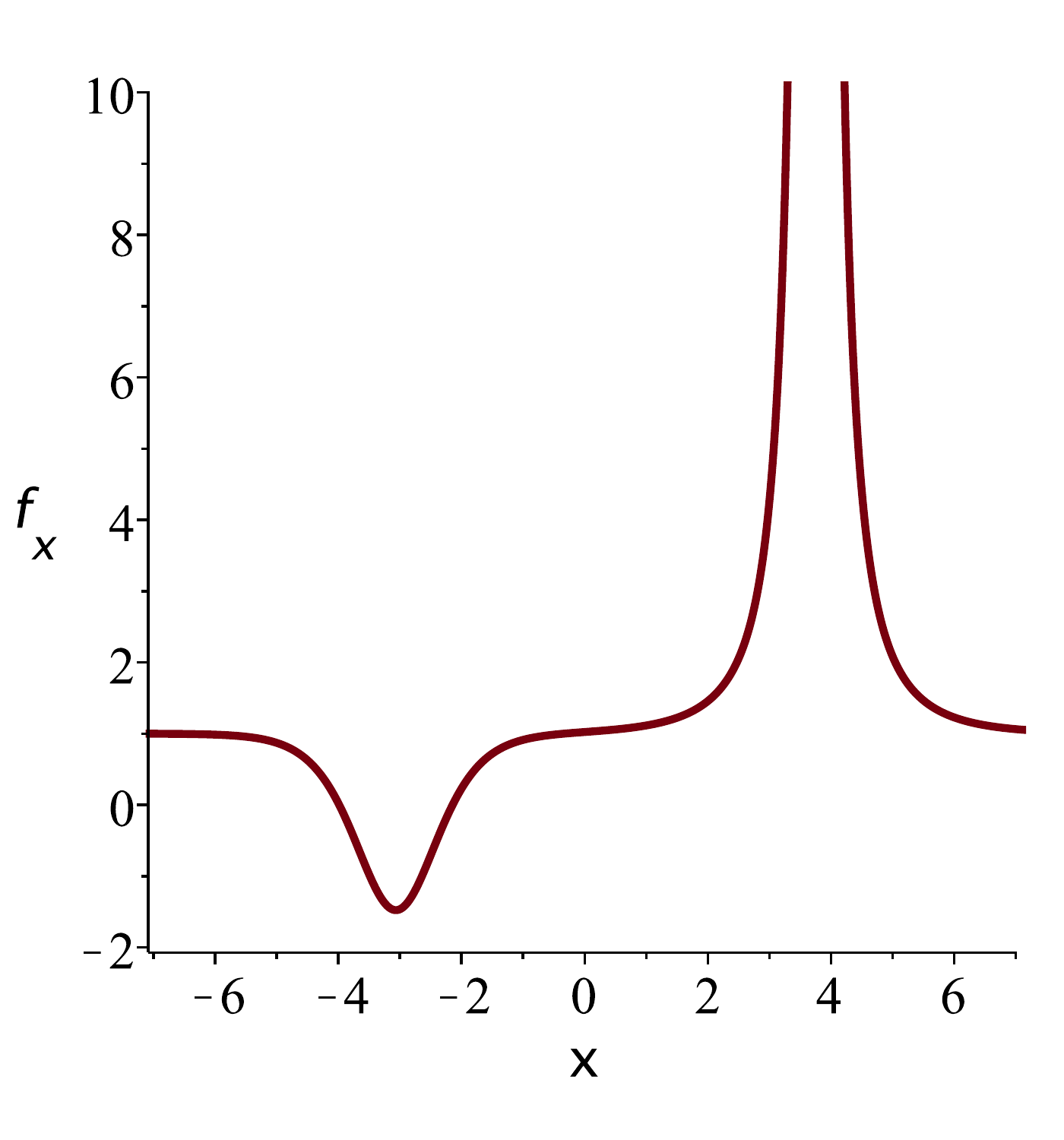} 
        }
      \caption{The expulsion of soliton by a singular soliton.  Parameter values are $\beta=1$ and $\theta=-1$, so $\lambda_1,\lambda_2,\lambda_3 \approx -1.88,0.35,1.53$, and
         $C_1=C_2=1,C_3=-1$. Plots of $f_x$,  displayed for times $t=-3,0,0.2,0.5,1,2$.}  
\end{figure}

\begin{figure}
      \centerline{
        \includegraphics[width=4.5cm]{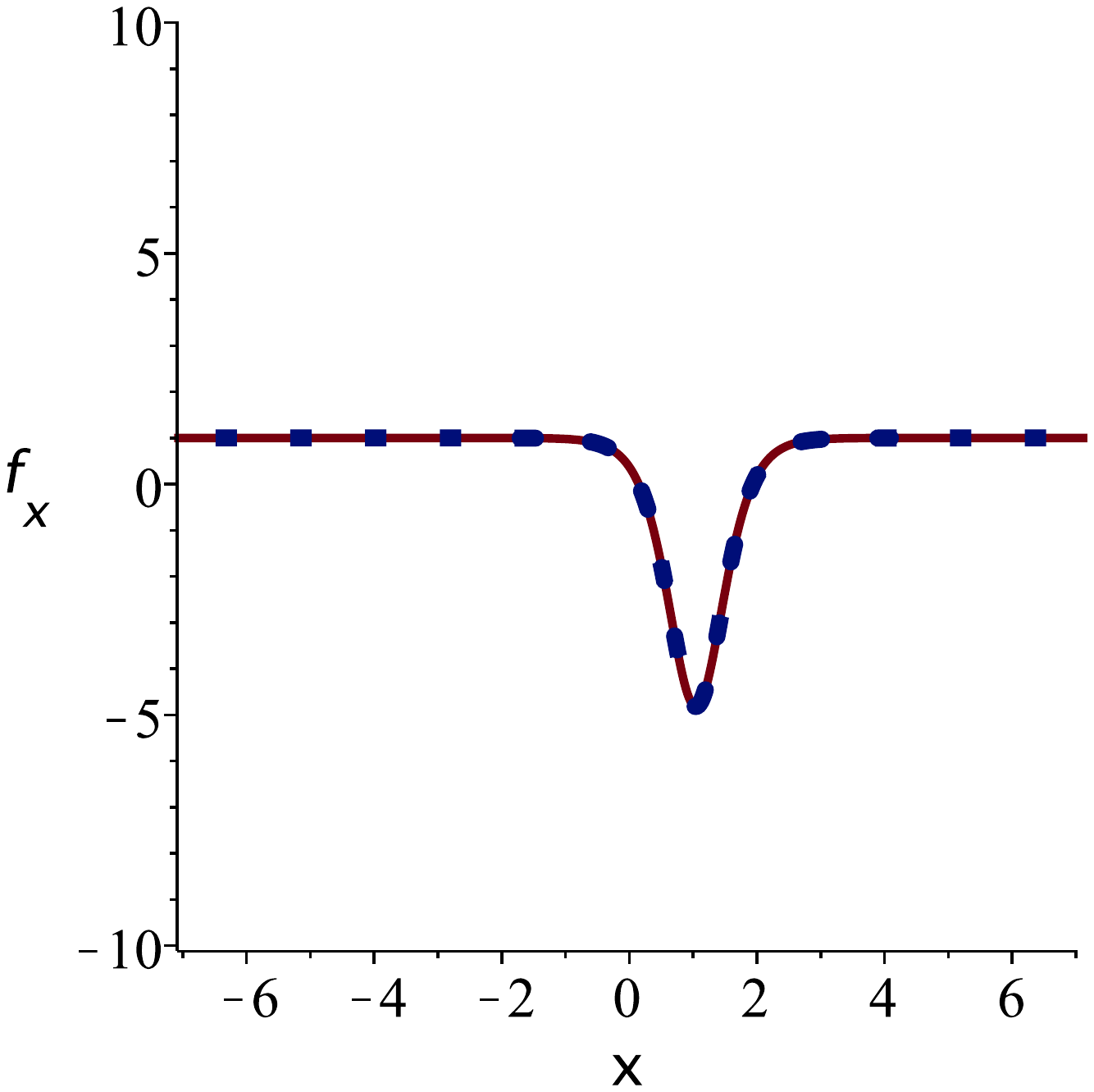} 
        \includegraphics[width=4.5cm]{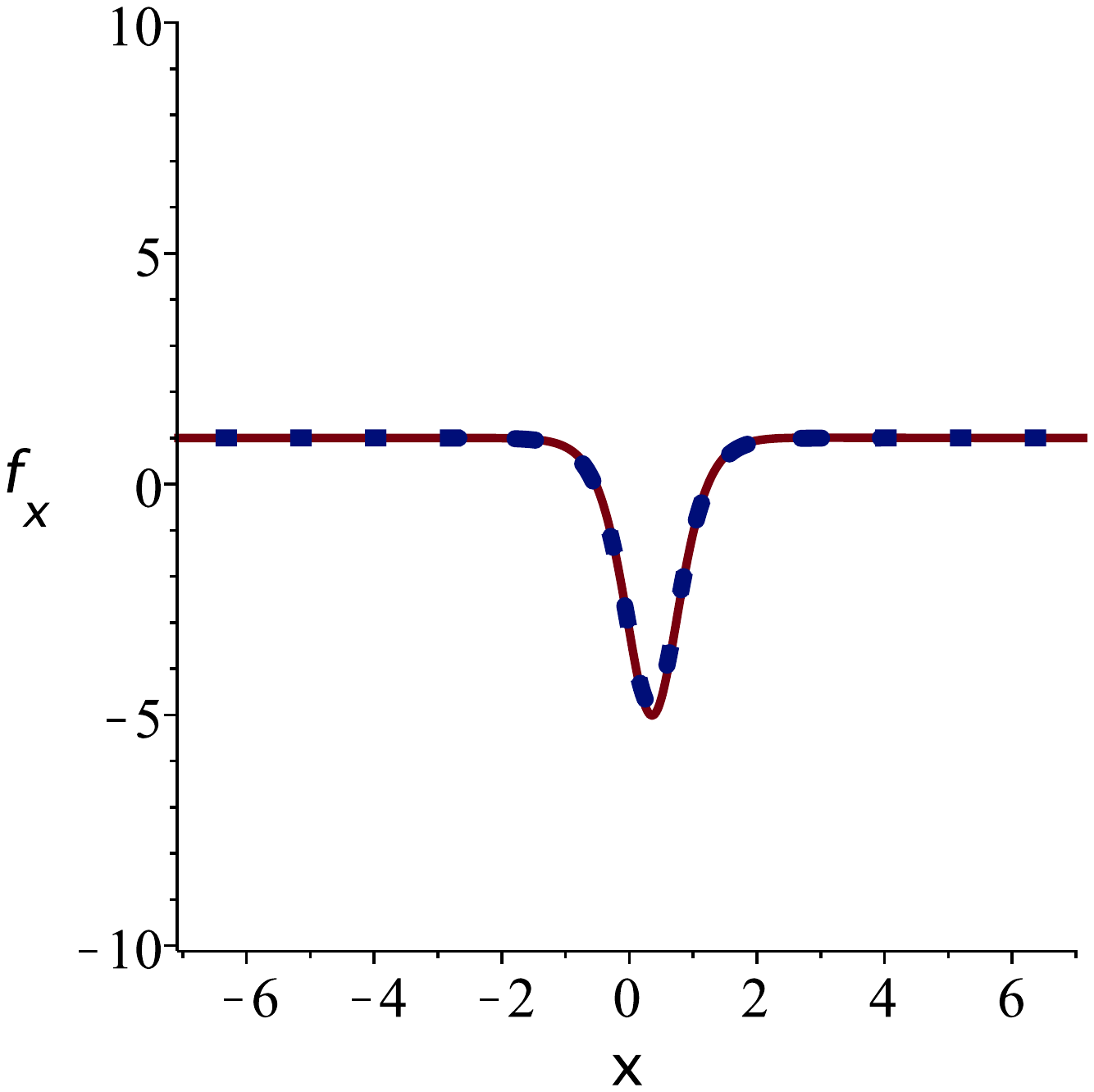} 
        \includegraphics[width=4.5cm]{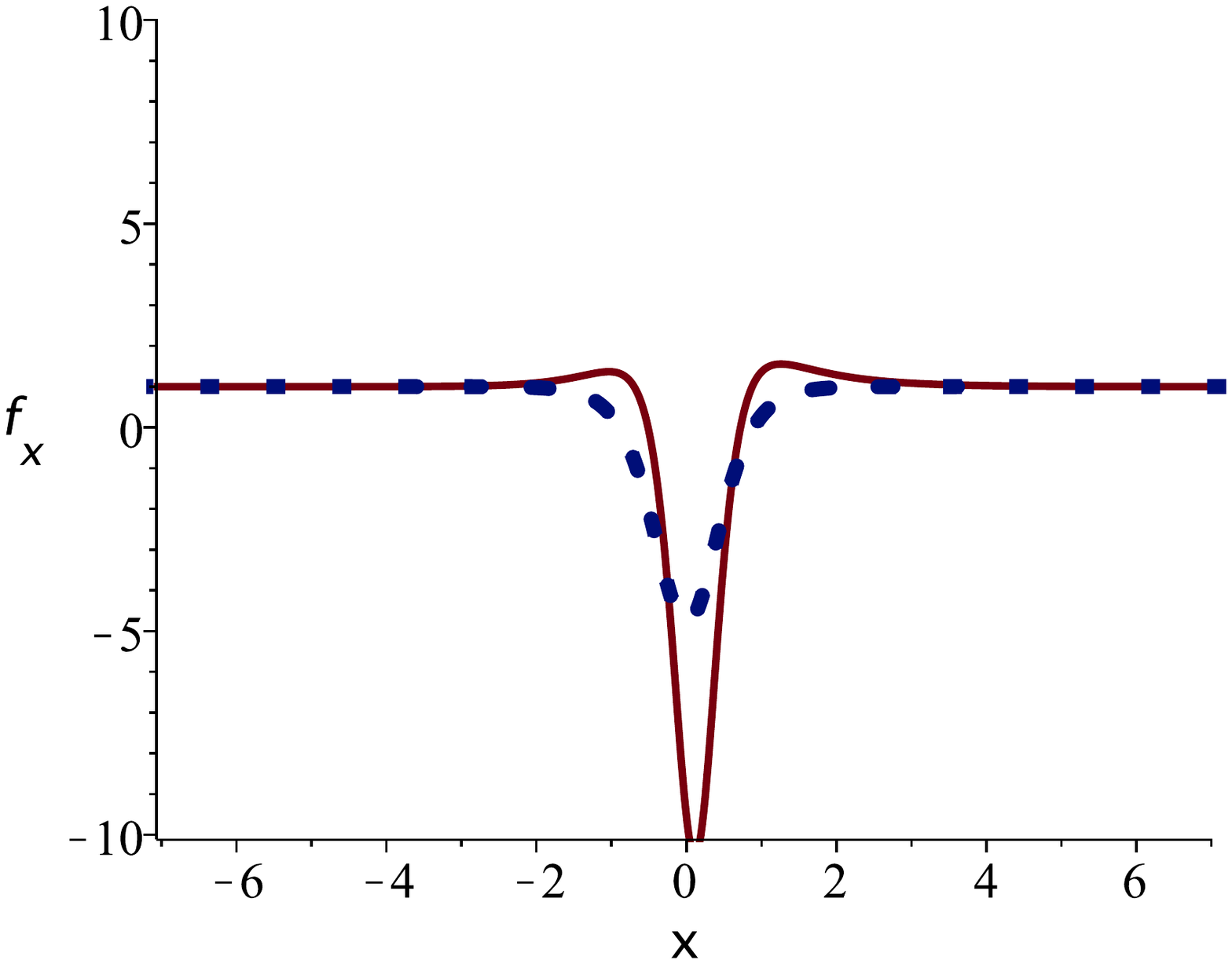} 
        }
      \centerline{
        \includegraphics[width=4.5cm]{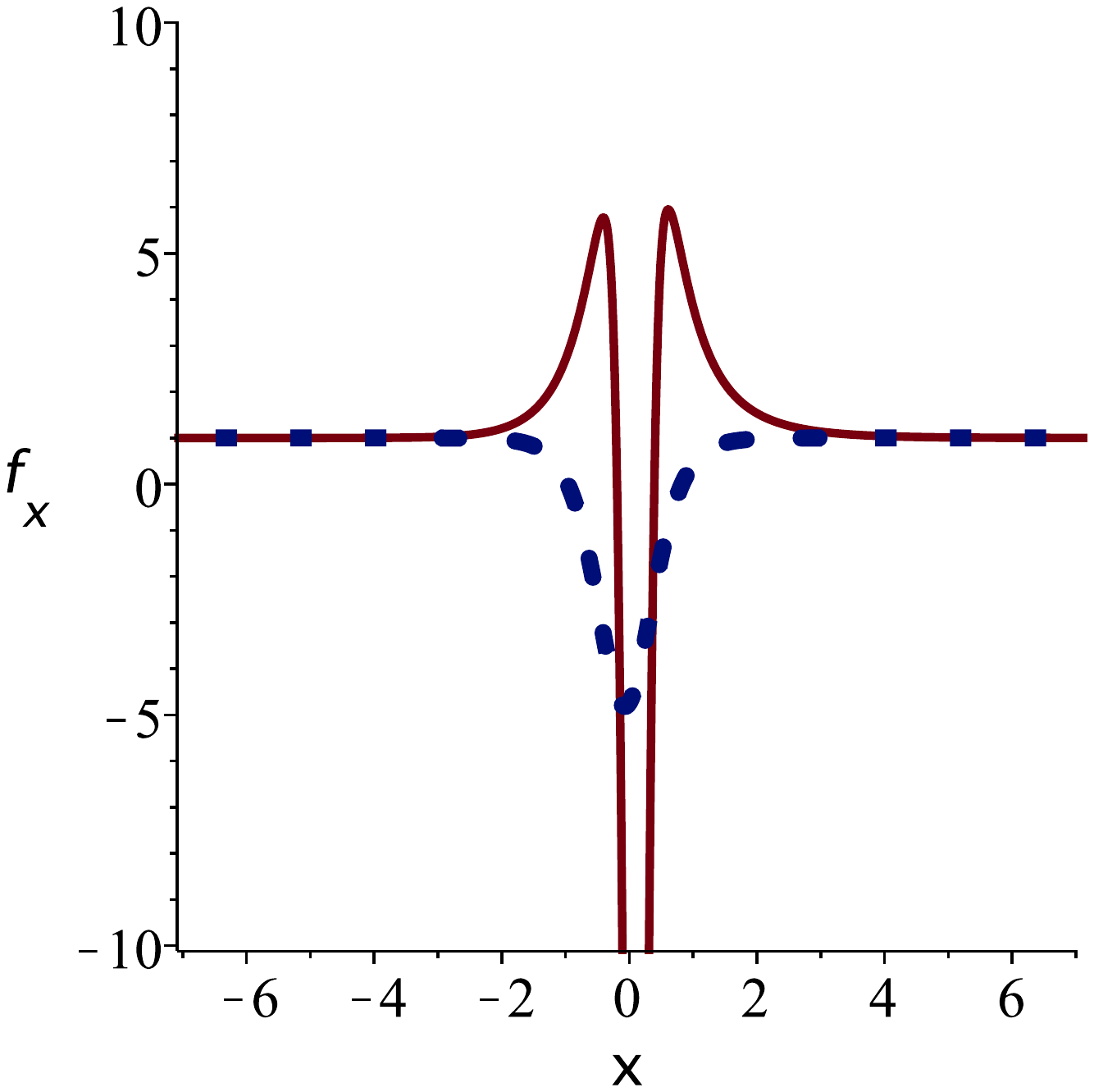} 
        \includegraphics[width=4.5cm]{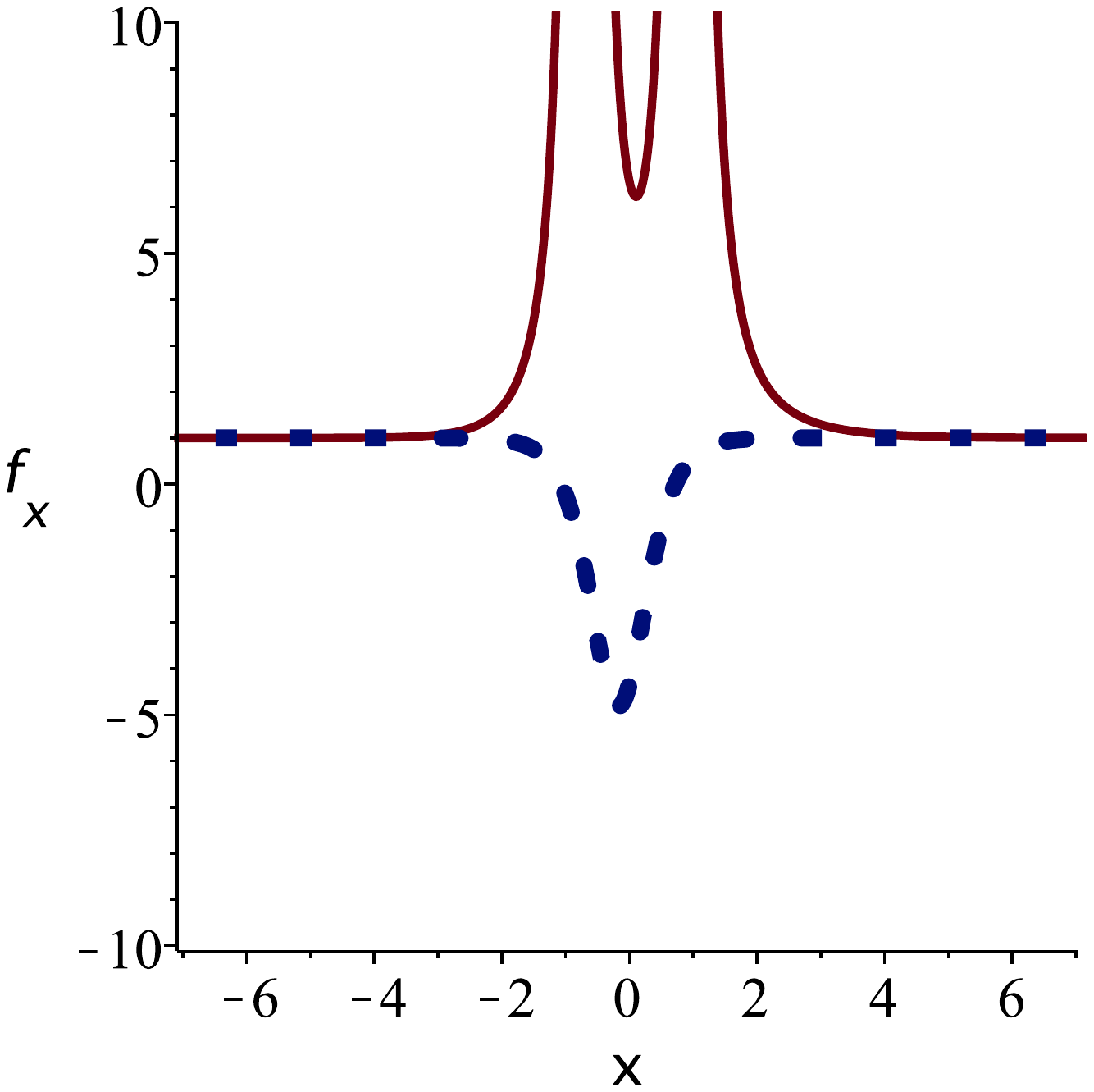} 
        \includegraphics[width=4.5cm]{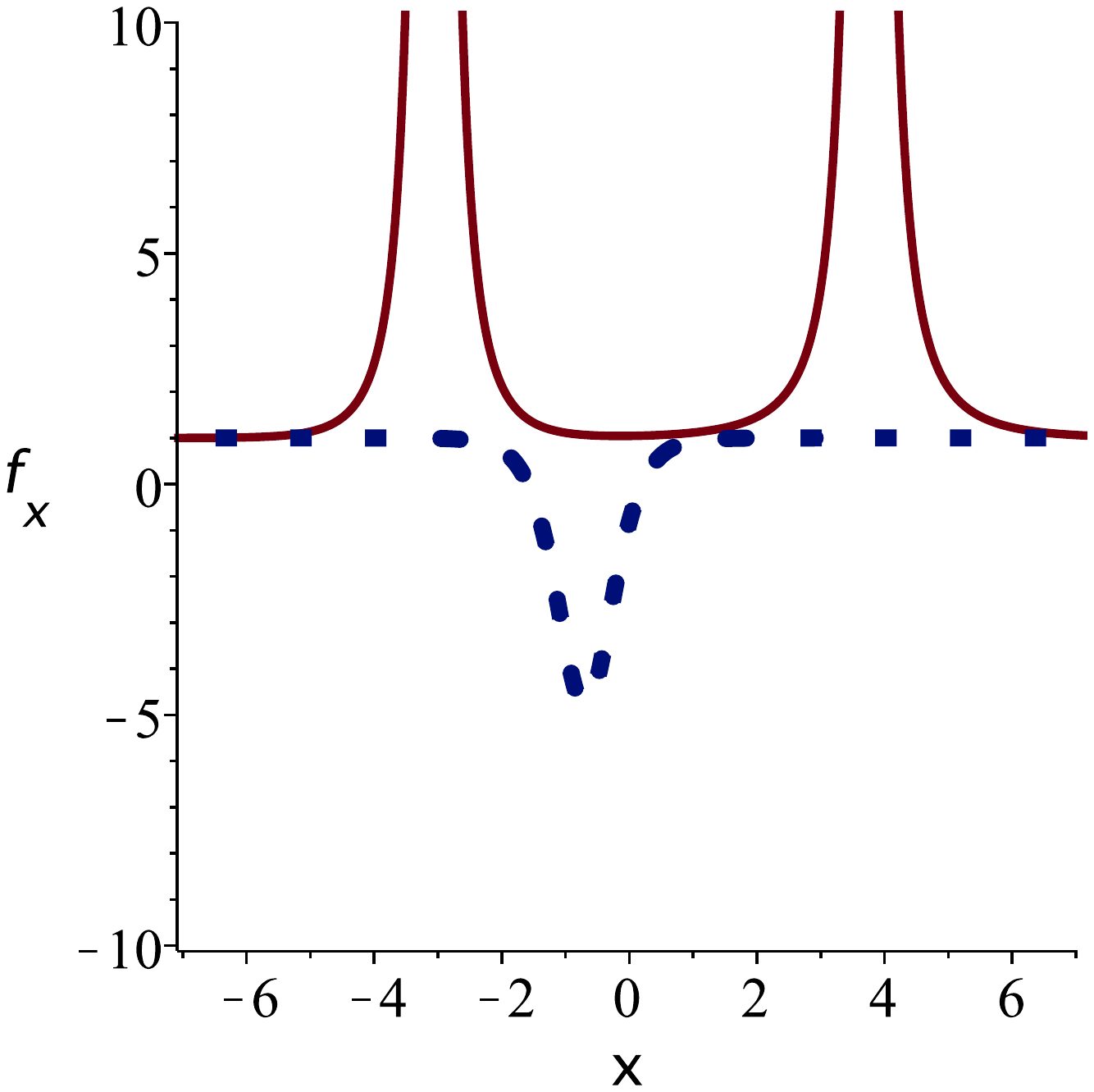} 
        }
      \caption{The splitting of a perturbed soliton into two singular solitons.  Parameter values $\beta=1$ and $\theta=-1$, so $\lambda_1,\lambda_2,\lambda_3
        \approx -1.88,0.35,1.53$, and  $C_1=C_3=1,C_2=-1$.
        Plots of $f_x$,  displayed for times $t=-3,-1,0,0.2,0.5,2$.
        The blue, dotted plots show the unperturbed soliton solution, given by the same parameters except $C_2=0$.} 
\end{figure}

\section{Conservation laws}
To find the conservation laws of aN, it is just necessary to observe that 
(\ref{cl}) has the form of a conservation law.
Thus $v$,  which is the solution of (\ref{vf1})-(\ref{vf2}) and depends on $\theta$,
provides a generating function for (densities of) conservation laws. Similar to what was  observed for the Camassa-Holm equation in
\cite{FS0}, this can be expanded in various different ways to obtain series of conservation laws. 

For large $|\theta|$,  the solution $v$ to (\ref{vf1}) can be expanded in an asymptotic series of the form
\begin{equation}
v \sim  \sum_{i=-1}^{\infty}\theta^{-i/3}v_i \ .\label{asymp} 
\end{equation}
Each of the coefficients $v_i$ is the density for a conservation law. 
The first few coefficients are given as follows: 
$$
v_{-1}^3 = 1 \ , \quad
v_0=0\ ,\quad
v_1=\frac{f_x}{v_{-1}}\ , \quad 
v_2=-\frac{f_{xx}}{v_{-1}^2}\ , \quad  
v_3=\frac{2}{3}f_{xxx}\ , \quad
v_4=-\frac{f_{xxxx}}{3v_{-1}}\ .
$$
Further terms can be computed using the recurrence relation
$$
v_{k+2} =\frac{1}{3v_{-1}^2}
   \left(
   3f_xv_{k}
  - \sum _{j=-1}^{k+1} \left( \sum_{i=\max(-1,-j-1)}^{\min(k+1,k-j+1)}v_{k-i-j}v_{i}v_{j} \right)
  -3\sum _{i=-1}^{k+1}v_{i}v_{(k-i)x}
  -v_{kxx} 
  \right)\ ,\quad k\geq1.
$$
So, for example 
\begin{eqnarray*}
v_5&=&\frac{f_{5x}+3f_xf_{xxx}-3f_x^3}{9v_{-1}^2}\ , \\
v_6&=&2f_x^2f_{xx}-\frac{2}{3}f_{xx}f_{xxx}-\frac{2}{3}f_xf_{xxxx} \ , \\
v_7&=&\frac{1}{27v_{-1}}\left(21f_{xxx}^2-72f_{x}^2f_{xxx}-117f_{x}f_{xx}^2+39f_{xx}f_{xxxx}+21f_{x}f_{5x}+9f_{x}^4-f_{7x}\right)\ .
\end{eqnarray*}
Here $f_{5x},f_{6x},\ldots$ denote 5th, 6th etc. derivatives of $f$ with respect to $x$. 
For each $i=1,2,\ldots$ $v_i$,  is the density $F$ of a conservation law $F_t + G_x = 0$. 
For $i=1,2,3,4,6$ the conservation laws are evidently trivial ($F=H_x$, $G=-H_t$ for some $H$). 
For $i=5,7$ we obtain the following conservation laws (after eliminating trivial parts): 
\begin{align*}
&F_5=f_xf_{xxx}-f_x^3,&~~~&F_7=24f_x^2f_{xxx}-3f_x^4+39f_xf_{xx}^2+f_{xx}f_{xxxx}, \\
&G_5=-(3f_x+f_{xx}f_{tx}),&~~~&G_7=3(18f_x^2-20f_tf_x^3-f_tf_xf_{xxx}-15f_{tx}f_xf_{xx}+f_{xxx})\ .
\end{align*}
Note that there are $3$ possible series for $v$, corresponding to the $3$ possible choices of $v_{-1}$.
The dependence of $v_1,v_2,\ldots$ on the choice of $v_{-1}$ is clear, and can be verified to be
consistent with the recursion relation. We denote the  three associated asymptotic series by $v^{(1)},v^{(2)},v^{(3)}$.
If we  define $\sigma =  v^{(1)} + v^{(2)} + v^{(3)}$, then $\sigma$ has 
asymptotic series $\sum_{i=1}^\infty 3v_{3i} \theta^{-i} $.  However, if we define 
\begin{equation}
A=(v^{(2)}-v^{(3)})v^{(1)}_x+(v^{(3)}-v^{(1)})v^{(2)}_x+(v^{(1)}-v^{(2)})v^{(3)}_x
  - (v^{(1)}-v^{(2)}) (v^{(2)}-v^{(3)})(v^{(3)}-v^{(1)})\ , 
\label{Aref}\end{equation}
then it can be verified (using (\ref{vf1}) for each of the functions $v^{(1)},v^{(2)},v^{(3)}$)  that
\begin{equation}
\sigma = v^{(1)}+v^{(2)}+v^{(3)} = -(\log A)_x \ . 
\label{Aid}\end{equation}
It follows that $v_{3i}$ is a total $x$ derivative for all $i$, and the associated
conservation laws are trivial. We also conjecture that $v_{2i}$ is a total $x$ derivative for all $i$, though
do not yet have a direct proof of this. 

Another possible expansion of $v$, this time for small $|\theta|$, is  
\begin{equation}
v \sim  \sum_{i=1}^{\infty}\theta^{i}w_i \ . \label{texp}
\end{equation}
The coefficients of this series are obtained by plugging (\ref{texp}) into (\ref{vf2}), and involve 
derivatives with the respect to $t$, which can not be eliminated. 
We call it the ``expansion in the $t$ direction'', as opposed to the (\ref{asymp}) which is the ``expansion in the $x$ direction''. 
The first few coefficients of (\ref{texp}) are 
$$
w_1=-\frac{1}{3}f_t\ , \quad 
w_2=-\frac{1}{9}f_{tt}\ , \quad  
w_3=\frac{1}{81}(f_t^2f_{ttx}-f_t^4-f_tf_{tt}f_{tx}-3f_{ttt})\ .
$$
For each $i=1,2,\ldots$, $w_i$ is the density of a conservation law.
The conservation law arising from $w_2$ is trivial, but from $w_1,w_3$, after elimination of some  trivial parts, we obtain the following
conservation laws: 
\begin{align*}
&\bar{F}_1=-3f_t,&~~~ &\bar{F}_3=-3f_{t}(f_{t}^3+3f_{tt}f_{tx}),\\
&\bar{G}_1=f_tf_{ttx}-f_t^3-f_{tx}f_{tt},&~~~&\bar{G}_3=15f_{t}f_{tt}^2-3f_{tt}f_{tttx}+3f_{ttt}f_{ttx}-2(f_t^3-f_tf_{ttx}+f_{tt}f_{tx})^2.
\end{align*}

\section{Symmetries}
By the direct computation of the first order generalized symmetries \cite{Ol0} of aN we obtain
\begin{eqnarray*}
  \eta^{(f)} &=&1 \ , \\
  \eta^{(x)} &=&f_x \ , \\
  \eta^{(t)} &=&f_t \ , \\
  \eta^{(s)} &=&xf_x-3tf_t+f\ ,
\end{eqnarray*}
corresponding to invariance of aN under translations of $f,x,t$ and a scaling symmetry. 

Following the ideas of \cite{RS2}, and with some inspiration from formulas that appeared in \cite{RS6,RS7},
it is possible to find a generating function for symmetries of aN in terms of mulitple solutions of the BT
(\ref{vf1})-(\ref{vf2}):
\begin{equation}   
Q = \frac{1}{A}((v^{(1)}v^{(2)}_x-v^{(1)}_xv^{(2)})v^{(3)}-(v^{(1)}-v^{(2)})(v^{(1)}v^{(2)}v^{(3)}+\theta))\ .   \label{gensym} 
\end{equation}
Here $A$ is defined by (\ref{Aref}).  It is straightforward to check directly that
$\eta = Q$ satisfies the linearized aN equation
\begin{equation}   
  \eta_{xxt} 
      -  3( f_x \eta_t + f_t \eta_x )  = 0  \ .  
\end{equation}
To generate local symmetries for aN from (\ref{gensym}) we take
$v^{(1)},v^{(2)},v^{(3)}$ to be given by the 3 possible asymptotic expansions of $v$ of form (\ref{asymp}), as
given in the previous section. Substitution of these asymptotic expansions
into $Q$ and expansion in inverse powers of $\theta^{1/3}$ gives an  infinite hierarchy of symmetries. The first two symmetries
in this hierarchy are $\eta^{(f)}, \eta^{(x)}$. From the higher order terms we obtain 
\begin{eqnarray*}
  \eta^{(5)} &=&\frac{1}{15}f_{5x}+f_x^3-f_xf_{xxx} \ , \\
  \eta^{(7)} &=&\frac{1}{21}f_{7x}-3f_x^4+6f_x^2f_{xxx}+3f_xf_{xx}^2-f_xf_{5x}-f_{xx}f_{xxxx}-f_{xxx}^2\ . 
\end{eqnarray*}
Observe that $\eta^{(5)}$ defines the potential Sawada-Kotera flow (\ref{pSK}). 

We have not succeeded in using the second expansion (\ref{texp}) of $v$ in (\ref{gensym}) to generate a second
hierarchy of symmetries, as the coefficients in this expansion are unique. However, we note that there are
other equations for which such an approach is possible. Specifically, for the Camassa-Holm equation 
\begin{equation}
m_t+2u_xm+um_x=0,\qquad    m=u-u_{xx},  \label{CM0}
\end{equation}
the system of the BT for the Camassa-Holm equation is
\begin{align}
s_{x}=&-\frac{s^2}{2\alpha}+\frac{1}{2}(m+\alpha)\ ,\label{BT1}\\
s_{t}=&-s^2\left(1-\frac{u}{2\alpha}\right)-u_xs+\frac{1}{2}(2\alpha^2+\alpha u-um)\ .\label{BT2}
\end{align}
One possible expansion of $s$, the ``expansion in the $x$ direction'',  is in the form
\begin{equation}
s=\sum_{n=1}^{\infty}s_n\alpha^{n/2} \ . \label{exp0}
\end{equation}
There are two versions of this expansion (related by replacing $\alpha^{1/2}$ by $-\alpha^{1/2}$), and these
were used in \cite{RS5} to construct a hierarchy of symmetries. 
However there is also an ``expansion in the $t$ direction'': 
\begin{equation}
s=\sum_{n=0}^{\infty}s_n\alpha^{1-n}\ .  \label{exp1}
\end{equation}
The first few coefficients are 
\[
s_0^2=1,~~s_1=\frac{1}{2s_0}(u-s_0u_x),~~s_2=\frac{1}{8s_0}(2u_{tx}-2s_0uu_x-u^2+2uu_{xx}-2s_0u_t+u_x^2).
\]
Once again, there are two versions of this expansion, corresponding to the choices $s_0=\pm 1$. 
A generating symmetry for the Camassa-Holm equation was found in \cite{RS5} and its form is
\begin{equation}
Q_{ch}=\frac{s^{(1)}+s^{(2)}}{s^{(2)}-s^{(1)}}\ .\label{symm1}
\end{equation}
In \cite{RS5} the two expansions of $s$ of form (\ref{exp0}) were used in this formula to produce a hierarchy of symmetries. 
A second hierarchy is found by using the two expansions of form (\ref{exp1}).  The first few symmetries  we get are
\begin{eqnarray*}
\eta^{(1)}&=&u_x\ ,  \\
\eta^{(2)}&=&u_t\ ,  \\
\eta^{(3)}&=&\left(2uu_{xt}-2u_{tt}+uu_x^2-2u_tu_x-5u^3+4u^2u_{xx}\right)_x,\\   
\eta^{(4)}&=&8u_xu^3-2u_{ttt}-2u_{xxx}u^3-6u_{x}u_{xx}u^2-3u_{t}(3u^2-2uu_{xx}-u_{x}^2)\ .
\end{eqnarray*} 
This hierarchy of symmetries is different from the one  found in \cite{RS5}.
We note that this is a hierarchy of hyperbolic, not evolutionary,  symmetries.
These symmetries have already been found, for example, in \cite{GKKV}.
The existence of hierarchies of such symmetries has been also discussed in \cite{AS0}.
In parallel to the fact that we have been able to find two hierarchies of conservation laws for aN,
we expect there to be two hierarchies of symmetries. But finding the second hierarchy remains an open problem. 

\section{Painlev\'e Property}

It is lengthy, but straightforward,  to verify that aN has the WTC Painlev\'e property \cite{WTC}. There are formal series solutions
about the singularity manifold $\phi(x,t)=0$  of the form
$$  f = \phi^{-1} \sum_{j=0}^\infty f_j \phi^j   $$
with $\phi,f_1,f_6$ arbitrary, except that  $\phi_x$ and  $\phi_t$ must not vanish. As usual, 
the truncated Painlev\'e series (see \cite{weiss1992}), which has the simple form  
$$ f = f_1 - \frac{2\phi_x}{\phi}\ ,  $$
gives the BT, c.f. equations (\ref{BT}) and (\ref{LP2BT}).   

\section{Scaling Reduction}

aN has the obvious scaling invariance $f\rightarrow \lambda f$,  $\partial_x \rightarrow \lambda \partial_x$,  $\partial_t \rightarrow \lambda^{-3} \partial_t$.
Thus we look for a reduction to an ordinary differential equation by taking
$$ f(x,t) =  \frac{F(z)}{3x} ,  \qquad   z  = 3xt^{1/3}  \ .  $$ 
This gives the third order equation
\begin{equation}
  F''' =   \frac{z{F'}^2-FF'}{z^2} - 1  \ .
  \label{toe} \end{equation} 
This equation has the Painlev\'e property, with a pole series
$$ F = \sum_{j=0}^\infty F_j(z-z_0)^{j-1}  $$
with $z_0,F_1,F_6$ arbitrary, except that $z_0\not=0$. 
The equation has a first integral
\begin{equation}
  E =  zF''^2 - F'F'' - \frac23 F'^3 + \frac{FF'^2}{z} + 2 z F' - 3 F\ .  \label{sd} 
\end{equation}  
Note that if $F$ satisfies the third order equation (\ref{toe}) then $E$ is constant. But in the other direction, 
$E$ being constant implies that either $F$ satisfies the third order equation (\ref{toe}) or
$F = C_1 + C_2 z^{3/2}$ where $C_1,C_2$ are constants. Remarkably, the functions
$$ F = \frac14 \pm \frac{2}{\sqrt{3}}z^{3/2}   $$
also satisfy the third order equation (\ref{toe}), and illustrate the fact that although
(\ref{toe}) has the Painlev\'e property, solutions can have a non-pole singularity at $z=0$. 
These give rise to the solutions $f = \frac1{12x} \pm 2 \sqrt{xt}$  of aN. 

By the substitution
$$  F(z) = \frac14 -  9 G(w) \  ,  \qquad   w = z^{3/2} \ , $$
equation (\ref{sd}) is brought to the form
$$  w^2 G''(w)^2 = - 4 \left (wG'(w) - G(w) \right) G'(w)^2  + \frac{16}{243}  \left (wG'(w) - G(w) \right)  + \frac{4(4E+3)}{6561}\ .  $$
This is a special case of equation SD-I.b, equation (5.5) in \cite{CS93}, which can be solved, as explained in \cite{CS93},  in terms
of either Painlev\'e III or Painlev\'e V transcendents. 

\section{Hirota form}

Substituting
$$ f = -2\frac{\phi_x}{\phi}  $$
in aN we obtain the Hirota bilinear form
$$ ( D_x^3 D_t - 3)  \phi\cdot \phi = 0  \ .   $$ 
See, for example, \cite{Hie0} for notation.  This has an unusual form, in that as far as we are aware, 
most studies of equations in Hirota form assume the form $P(D_x,D_t,\ldots) \phi\cdot\phi = 0 $
with $P(0,0,\ldots)=0$ (see, for example, \cite{Hie0}, equation (17)). However, by substituting 
$$ \phi = C(t) e^{-\beta x^2/4 - t x /2 \beta}  \psi \ , $$
where $C(t)$ is an arbitrary function,  we obtain
$$ \left( D_x^3 D_t + \frac{6}{\beta} D_x^2 + 6\beta D_xD_t  \right)  \psi\cdot\psi  = 0   $$ 
Modulo rescalings, this is the Hirota form of the Hirota-Satsuma equation  \cite{Hie0,Hie1,ma1990}. 
The basic soliton solution is
$$  \psi = 1 +  e^{\eta} \ , \qquad  \eta = a( x - ct) +  b  $$
where $a,b,c$ are constants with
$$   a^2  =  3 \left( \beta - \frac{1}{\beta c}  \right)\ .    $$ 
As observed in our previous discussion of soliton solutions, the speed $c$ is restricted by the requirement
$\frac{1}{\beta c} < \beta$.
The  two soliton solution is
$$
\psi = 1 +  e^{\eta_1} + e^{\eta_2} + A e^{\eta_1+\eta_2}  
$$
where
\begin{eqnarray*}
  \eta_1 &=&   a_1 ( x - c_1 t)   + b_1 \   , \qquad    a_1^2  =  3 \left( \beta - \frac{1}{\beta c_1} \right) \ , \\
  \eta_2 &=&   a_2 (x - c_2 t)    + b_2 \   , \qquad    a_2^2  =  3 \left( \beta - \frac{1}{\beta c_2} \right) \ , \\
%  A &=&   \frac
%  {\left( 3c_1 c_2(c_1+c_2)\beta^2 -(c_1^2+4c_1c_2+c_2^2) \right) \beta a_1a_2  -9(c_1+c_2)\left( \beta^4  c_1 c_2  -  \beta^2 (c_1+c_2) + 1   \right)  }
%  {\left( 3c_1 c_2(c_1+c_2)\beta^2 -(c_1^2+4c_1c_2+c_2^2) \right) \beta a_1a_2  +9(c_1+c_2)\left( \beta^4  c_1 c_2  -  \beta^2 (c_1+c_2) + 1   \right)  }
  A &=&  \frac
  {  (3c_1c_2(c_1+c_2)\beta^2- (c_1^2+4c_1c_2+c_2^2) - \beta c_1c_2a_1a_2(c_1+c_2) )^2 }
  {(c_1-c_2)^2 (3\beta^2c_1c_2(c_1+c_2) + c_1^2+c_1c_2 + c_2^2)  } \ . 
\end{eqnarray*}
Here we have written the phase shift factor $A$ in a form in which it is clear that at least in the case that both $c_1$ and $c_2$ are positive,
$A$ is positive, thus guaranteeing a non-singular two soliton solution of aN.  Note that for certain values of the parameters $A$ can either vanish,
or be ill-defined, as the denominator vanishes.  It can be checked that this happens  if and only if 
$$  \frac{c_1^3}{1+3\beta^2 c_1} = \frac{c_2^3}{1+3\beta^2 c_2}  $$
In this case the two soliton solution reduces to  the ``merging soliton'' solutions  described in section 3.
The two soliton solution extends to multisoliton solutions in the usual way for integrable equations of KdV type \cite{Hie0,Hie1}. 
So, for example, the three soliton solution is, in the obvious notation, 
$$  
\psi = 1 +  e^{\eta_1} + e^{\eta_2} + e^{\eta_3} +  A_{12} e^{\eta_1+\eta_2}  +  A_{13} e^{\eta_1+\eta_3}  +  A_{23} e^{\eta_2+\eta_3}  + A_{12}A_{13}A_{23} e^{\eta_1+\eta_2+\eta_3} \ .  
$$
We have not yet succeeded in determining whether there exist nonsingular superpositions of merging solitons and regular solitons. 

\section{Conclusion} 
In this paper we have researched the aN equation, an integrable nonlinear equation, which is notable for taking a particularly simple form,
with just a single, quadratic nonlinear term,  and for having relationships with various other significant  integrable equations.
Among the interesting properties of this equation we would mention the existence
of merging solitons, the fact that some of the soliton solutions are unstable, which can be demonstrated very effectively, and the existence
of two infinite hierarchies of conserved quantities. Among the matters that we have not been able to resolve fully and which merit further research,
we would mention the lack of a superposition principle for the B\"acklund transformation, the need to establish the stability of other solutions,
and the need for a fuller exploration of the space of solutions, particularly to determine if there are nonsingular ways to superpose merging solitons. 
Also, we have only been able to give a single hierarchy of local symmetries so far. We note that the aN equation is not in evolutionary form,
which limits its potential application;  but its superficial simplicity and rich properties make it notable, nonetheless. 

\bibliographystyle{ap-jnmp}   % link to ap-jnmp.bst
\bibliography{P} 
\end{document}